\documentclass[runningheads]{svmult}
\usepackage{makeidx}   
\usepackage{graphicx}  
\usepackage{subeqnar}  
\usepackage{physprbb}  
\def\etal{{\it et al.}\ }

\def\KK{\rm ~K}

\def\EE#1{\times 10^{#1}}

\def\ccm{\rm ~cm^{-3}}
\def\kms{\rm ~km~s^{-1}}
\def\ergs{\rm ~erg~s^{-1}}

\def\La{${\rm Ly}\alpha$}
\def\Ha{${\rm H}\alpha$}

\def\Msun{M_\odot}
\def\Msunyr{M_\odot~\rm yr^{-1}}
\def\Mu5{\dot{\cal{M}}}
\def\Mdot{\dot M}
\def\lsim{\!\!\!\phantom{\le}\smash{\buildrel{}\over
  {\lower2.5dd\hbox{$\buildrel{\lower2dd\hbox{$\displaystyle<$}}\over
                               \sim$}}}\,\,}
\def\gsim{\!\!\!\phantom{\ge}\smash{\buildrel{}\over
  {\lower2.5dd\hbox{$\buildrel{\lower2dd\hbox{$\displaystyle>$}}\over
                               \sim$}}}\,\,}
\begin{document}
\title*{Supernova Interaction with a Circumstellar Medium}
%
%
%
%
\titlerunning{Supernova Interaction with a Circumstellar Medium}
%
\author{Roger A. Chevalier\inst{1}
\and Claes Fransson\inst{2}}
\authorrunning{Chevalier and Fransson}
%
%
\institute{Department of Astronomy, University of Virginia, P.O. Box 3818, \\
Charlottesville, VA 22903, USA
\and Stockholm Observatory, Department of Astronomy,
SCFAB, \\
 SE-106 91 Stockholm, Sweden}

\maketitle              

\begin{abstract}
The explosion of a core collapse supernova drives a powerful shock
front into the wind from the progenitor star.
A layer of shocked circumstellar gas and ejecta develops that is
subject to hydrodynamic instabilities.
The hot gas can be observed directly by its X-ray emission, some
of which is absorbed and re-radiated at lower frequencies by the
ejecta and the circumstellar gas.
Synchrotron radiation from relativistic electrons accelerated at the
shock fronts provides information on the mass loss density if free-free
absorption dominates at early times or the size of the emitting region
if synchrotron self-absorption dominates.
Analysis of the interaction leads to information on the density and structure
of the ejecta and the circumstellar medium, and the abundances in
these media.
The emphasis here is on the physical processes related to the
interaction.
\end{abstract}

\section{Introduction}
\label{intro}
The collision of supernova ejecta with dense surrounding gas
can generate high pressure regions comparable to those found
in the emission regions of active galactic nuclei.
The shock wave interactions are analogous to those occurring
in $\gamma$-ray burst afterglows, although the supernova case
involves nonrelativistic velocities.
Multiwavelength observations, when combined with models for the
emission, give detailed information on the outer supernova structure
and the structure of the surrounding mass loss region.
The stellar mass loss history leading up to the supernova explosion
can be deduced from the surrounding structure.

The study of circumstellar interaction around supernovae has
benefitted from several developments in the observational investigations
of supernovae.
One is the ability to observe supernovae in wavelength regions other
than optical.
Radio, X-ray, infrared, and ultraviolet observations have all been
crucial for revealing various aspects of circumstellar interaction.
Another is the ability to follow the optical emission from supernovae
to late times with the new generation of large telescopes.
The emission from circumstellar interaction typically has a lower decay
rate than does the inner emission from gas heated by the initial
explosion and by radioactivity, so that it eventually comes to
dominate the emission.
Finally, the interest in supernovae generated by SN 1987A and other
events has increased the rate of supernova discovery, including
events which show signs of circumstellar interaction from a young age.

Many of the observational manifestations of circumstellar interaction
are treated in other chapters of this volume.
The emphasis here is on providing the theoretical basis to
model the observed phenomena.
The structure of the freely expanding supernova ejecta and
of the surrounding medium provide the initial conditions for the
interaction and are discussed in the first two sections.
Section 4 deals with the hydrodynamics of the interaction and section
5 treats the emission from hot gas.
The effects of supernova radiation on the circumstellar gas are discussed
in section 6
 and relativistic particles in section 7.
The discussion and conclusions are in
section 8.

\section{Ejecta Structure}
\label{ejecta}
The density structure of a supernova is set up during the first
days after the explosion.
Over this timescale, the pressure forces resulting from the
initial explosion, and some later power input from radioactivity,
become too small to change the supernova density structure.
The velocity profile tends toward that expected for free
expansion, $v=r/t$, and the density of a gas element drops
as $t^{-3}$.
The pressure drops to a sufficiently low value that it is not
a factor in considering shock waves propagating in the ejecta
as a result of later interaction.

The density structure of the ejecta is a more complex problem.
For core collapse supernovae, where all the explosion energy
is generated at the center of the star, the explosion physics
lead to an outer, steep power law density profile along with
an inner region with a relatively flat profile \cite{CS89,MM99}.
The outer profile is produced by the acceleration of the
supernova shock front through the outer stellar layers with
a rapidly decreasing density.
This part of the shock propagation does not depend on the
behavior in the lower layers and the limiting structure is
described by a self-similar solution.
The  structure depends on the initial structure of 
the star and thus depends on whether the progenitor star has
a radiative or a convective envelope \cite{MM99}.
In the radiative case, which applies to  Wolf-Rayet
stars and to progenitors like that of SN 1987A, the limiting
profile is $\rho\propto r^{-10.2}$.
In the convective case, which applies to red supergiant progenitors,
the limiting
profile is $\rho\propto r^{-11.7}$.
These are the limiting profiles and the density profile over
a considerable part of the supernova might be described by
a somewhat flatter profile.
Numerical calculations of the explosion of SN 1987A have indicated
$\rho\propto r^{-(8-9)}$ in the outer parts of the supernova \cite{A88}.

The calculations of the density profile assume an adiabatic flow.
As the shock wave approaches the surface of the star, radiative
diffusion becomes a factor and the shock acceleration stops.
In the case of SN 1987A, theory as well as observations suggest
that diffusion is not important until a velocity of
$\sim 30,000$ km s$^{-1}$ is reached \cite{EB92}.
A comparable limit applies to the relatively compact progenitors
of the Wolf-Rayet star explosions \cite{MM99}.
In the case of red supergiant progenitors, which is the case
applicable to most Type II supernovae, a lower velocity limit
is expected.
At the point in the shock evolution where radiative diffusion
becomes important, the loss of the radiative energy can lead to
a large compression of the gas, to the point where the thermal
pressure becomes important \cite{C76}.
There is the possibility of a dense shell at the maximum velocity
generated by the supernova.
The shell may be broken up by Rayleigh-Taylor instabilities
\cite{FA73} and the final outcome of these processes is not known.

The overall result of these considerations is that the outer part
of a core collapse supernova can be approximated by a steep power
law density profile, or $\rho_{\rm ej}\propto r^{-n}$ where $n$
is a constant.
After the first few days the outer parts of the
ejecta expand with constant velocity, $V(m)
\propto r$ for each mass element, $m$, so that $r(m) = V(m) t$ and $\rho(m) =
\rho_o(m) (t_o/t)^3$. Therefore
\begin{equation}
\rho_{\rm ej} = \rho_o
(t/t_o)^{-3} (V_o t/r)^{n}.  
\label{eq1a}
\end{equation}
This expression takes into account the free expansion of the gas.

Type Ia supernovae, which are believed to be the thermonuclear
explosions of white dwarfs, are different because the supernova
energy is not all centrally generated.
A burning front spreads through the star and the energy released
in this way gives rise to an explosion.
The results of a number of explosion simulations show that
the post-explosion density profile can be approximated by
an exponential in velocity \cite{DC98}.

\section{Stellar Mass Loss}
\label{mloss}
Type II supernovae, with H lines in their spectra, are thought
to be the explosions of massive stars that have reached the ends
of their lives with their H envelopes.
In most cases, the stars explode as red supergiants, which are
known to have slow, dense winds.
Typical parameters are a mass loss rate of $\dot M=10^{-6}-10^{-4}
~M_\odot ~\rm yr^{-1}$ and a wind velocity $u_{\rm w}=5-25 \rm ~km~s^{-1}$.
If the mass loss parameters stay approximately constant 
 leading up to the explosion, the circumstellar density
is given by
\begin{equation}
\rho_{\rm cs}=\dot M/(4\pi u_{\rm w} r^2).
\label{eq1b}
\end{equation}
In the late stellar evolutionary phases, the evolution of the
stellar core occurs on a rapid timescale, but the stellar envelope
has a relatively long dynamical time, which can stabilize the
mass loss properties.

The mechanisms by which mass is lost in the red supergiant phase
are poorly understood, but some insight into the wind properties can
be gained by considering observations of the winds.
VY CMa, thought to have a zero-age main sequence 
mass of $30-40~\Msun$ \cite{WLW98},
 a mass loss rate $\sim 3\times 10^{-4}~M_\odot~\rm yr^{-1}$, 
 and $u_{\rm w}\approx 
 39\kms$ \cite{De94,Se01}, is an especially well-observed case.
{\it HST} imaging shows that the density profile is approximately
$r^{-2}$ over the radius range $3\times 10^{16}-1.4\times 10^{17}$ cm
but that there is considerable structure superposed on this profile,
including knots and filaments \cite{Se01}.
Observations of  masers imply the presence of clumps with
densities up to $\sim 5\times 10^9$ cm$^{-3}$ and suggest the
presence of an expanding disk seen at an oblique angle \cite{RYC98}.
VY CMa is an extreme mass loss object, but there is evidence for
irregular mass loss in $\alpha$ Orionis and other red supergiants
\cite{THB97,UDG98}.
Among lower mass stars, the bipolar structure of planetary nebulae
has been widely attributed to a high equatorial density in
the red giant wind.
In one case where we can observe the circumstellar surroundings
of a supernova, SN 1987A, an axisymmetric structure similar to
that found in planetary nebulae has been observed \cite{Be95}.

Some massive stars, either through individual mass loss or through
binary interaction, lose their H envelopes entirely and become
Wolf-Rayet stars.
Their ultimate explosions are thought to be observed as Type Ib and
Ic supernovae.
The progenitor stars are more compact in this case and have faster
winds, with $u_w=1,000-2,500 \rm ~km~s^{-1}$ and $\dot M=10^{-6}-10^{-4}
~M_\odot ~\rm yr^{-1}$ \cite{Wi91}.
The fast wind can create a bubble in the surrounding medium, which is
typically the slow wind from a previous evolutionary phase, and
the resulting shells have been observed around a number of
Wolf-Rayet stars \cite{GS96}.
Their typical radii are a few pc.

The progenitors of Type Ia supernovae are not known, so that
observations of circumstellar interaction as well as other signatures of the companion star 
potentially contain
important clues on the progenitor question \cite{Br95,CL96,MBF00}.
In models where a white dwarf accretes mass from a companion star,
the wind from the companion can provide a relatively dense
circumstellar medium.
In the case of a double degenerate progenitor, there may be
a disk around the coalesced object, but the interaction is
with the interstellar medium on larger scales.
There have been no detections of circumstellar interaction around
Type Ia supernovae, so we do not discuss them further in this review,
although the physical processes described here should apply if any
surrounding wind is present.

\section{Hydrodynamics}
When the radiation dominated
shock front in a supernova nears the stellar surface,
a radiative precursor to the shock forms when the radiative
diffusion time is comparable to the propagation 
time.   There is radiative
acceleration of the gas and
the shock disappears when optical depth $\sim$unity is reached \cite{EB92}.
The fact that the velocity decreases with radius implies that
the shock will re-form as a viscous shock in the circumstellar
wind.
This occurs when the supernova has approximately doubled in radius.

The interaction of the ejecta, expanding with velocity $\gsim 10^4 \kms$, 
and the
nearly stationary circumstellar medium results in a reverse shock wave  
propagating inwards (in mass),
and an outgoing circumstellar shock. The
density in the circumstellar gas is given by equation (\ref{eq1b}).
 As discussed above, hydrodynamical calculations show that to a good 
 approximation
the  ejecta density can be described by equation (\ref{eq1a}).
A useful similarity solution for the interaction
can then be found \cite{C82a,C82b,N85}. 
Here we  sketch a simple derivation. More details can be found in
these papers, as well as in the review \cite{C90}. 

Assume that the shocked
gas can be treated as a thin shell with mass $M_{\rm s}$, 
velocity $V_{\rm s}$, and  radius $R_{\rm s}$. 
Balancing the ram pressure from the circumstellar   gas and the
impacting ejecta, the momentum equation for the shocked shell of 
circumstellar gas and
ejecta is
\begin{equation}   
M_{\rm s}{dV_{\rm s} \over dt}
= 4 \pi R_{\rm s}^2[\rho_{\rm ej} (V - V_{\rm s})^2 -
 \rho_{\rm cs}  V_{\rm s}^2] .
\label{eq2}
\end{equation}  
Here, $M_{\rm s}$ is the sum of the mass of the shocked ejecta and circumstellar   gas. The swept up
mass behind the circumstellar shock is $M_{\rm cs}=\Mdot R_{\rm s}/u_{\rm w}$, and that behind the
reverse  shock $M_{\rm rev}=4 \pi ~t_o^3 V_o^n (t/R_{\rm s})^{n-3}/(n-3)$, assuming that $R_{\rm s} >>
R_p$, the radius of the progenitor. With $V=R_{\rm s}/t$ we 
obtain  
\begin{eqnarray}
&&\left[{\Mdot \over u_{\rm w}} R_{\rm s} 
+ {4 \pi ~\rho_o~t_o^3 ~V_o^n~
t^{n-3} \over (n-3) ~R_{\rm s}^{n-3}} \right] {d^2 R_{\rm s} \over dt^2} = \nonumber \\
&&4 \pi R_{\rm s}^2 
\left[ 
{\rho_o ~t_o^3 ~V_o^n~t^{n-3} \over R_{\rm s}^n} 
\left({R_{\rm s} \over t} - {dR_{\rm s} \over dt}\right)^2 
- {\Mdot \over 4 \pi~u_{\rm w} R_{\rm s}^2}  \left({dR_{\rm s} \over
dt}\right)^2 \right].  
\label{eq3}
\end{eqnarray}
This equation has the power law solution
\begin{equation}
R_{\rm s}(t) = \left[{8 \pi \rho_o~t_o^3 ~V_o^n~u_{\rm w} \over (n-4) (n-3)
 ~\Mdot}\right]^{1/(n-2)} ~t^{(n-3)/(n-2)} .
\label{eq4}
\end{equation}
The form of this similarity solution can be written down directly by
dimensional analysis from
the only two independent quantities available, $\rho_o~t_o^3 ~V_o^n$ and
$\Mdot/u_{\rm w}$.  The solution applies after a few expansion times, when
the initial radius has been `forgotten.'  
The requirement of a finite energy in the flow implies $n > 5$.
More accurate
similarity solutions, taking  the structure within the shell into
account, are given in \cite{C82a,N85}. In general,
these solutions differ by less than $\sim 30 \%$ from the thin shell
approximation. 
 
\begin{figure}
\begin{center}
\includegraphics[width=12cm]{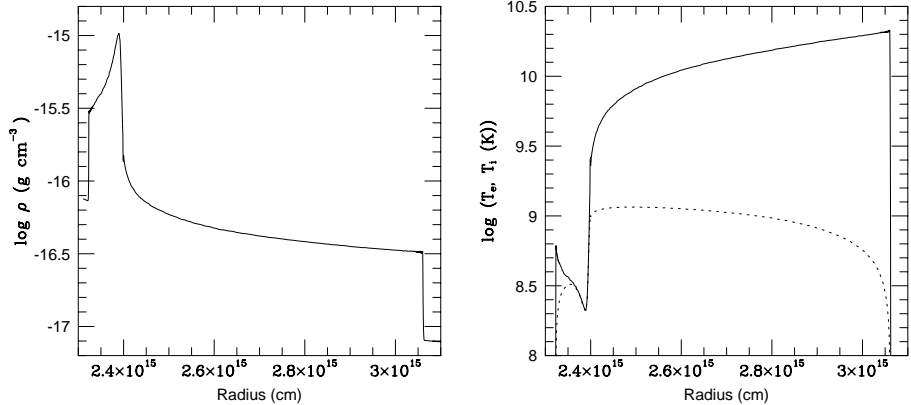}
\end{center}
\caption[]{Density and temperature structure of the reverse and circumstellar shocks for
$n=7$ and a velocity of $2.5\EE4 \kms$ at 10 days. Both shocks are assumed 
to be adiabatic. Because of the 
slow Coulomb equipartition the 
electron temperature (dotted line) is much lower than the ion temperature
 (solid line)
 behind the circumstellar shock.}  
\label{fig1}
\end{figure}

The maximum ejecta velocity close to the reverse shock depends on
time as $V = R_{\rm s}/t \propto t^{-1/(n-2)}$. 
 The velocity of the circumstellar shock, $dR_{\rm s}/dt$, in terms of $V$ is  
 $V_{\rm s} = V (n-3)/(n-2) $ and the reverse shock velocity, $V_{\rm rev}=  V -
V_{\rm s} = V/(n-2)$. Assuming cosmic abundances and equipartition between ions and electrons,
the  temperature of the shocked circumstellar gas is
\begin{equation}
T_{\rm cs} =
1.36\EE9 ~Ê\left({n-3\over n-2}\right)^2 \left(V\over 10^4 \kms \right)^2 
 ~\rm K
\label{eq5}
\end{equation}
and at the reverse shock
\begin{equation}
T_{\rm rev}= {T_{\rm cs}\over (n-3)^2}.
\label{eq6}
\end{equation}
The time scale for
equipartition between electrons and ions is 
\begin{equation}
t_{\rm eq} \approx 2.5\EE7 ~ \left({T_{\rm e} \over 10^9
\rm ~ K}\right)^{1.5}~  \left({n_{\rm e} \over 10^7 {\rm~cm}^{-3}}\right)^{-1} \rm~ s.
\label{eq6b}
\end{equation}
One finds that the
reverse shock is  marginally in equipartition, unless the temperature is $\gsim 5\EE8$
K. The ion temperature behind the
circumstellar shock is $\gsim 6 \EE{9}$ K for $V_4 \gsim 1.5$, and the
density a factor $\gsim 4$ lower than behind the reverse shock. 
Ion-electron collisions are therefore ineffective, and $T_{\rm e} <<
T_{\rm ion}$, unless efficient plasma instabilities heat the
electrons collisionlessly (fig. \ref{fig1}).

For typical parameters,
the electron temperatures of the two shocks are
very different,  $\sim (1-3)\EE9 \KK$ for the circumstellar shock and
$10^7 - 5\EE8 \KK$ for the reverse shock, depending on $n$. 
The radiation from 
the reverse shock is 
mainly below $\sim 20$ keV, while that from the circumstellar shock 
is above $\sim 50
\rm~keV$.  

\begin{figure}
\begin{center}
\includegraphics[width=9cm]{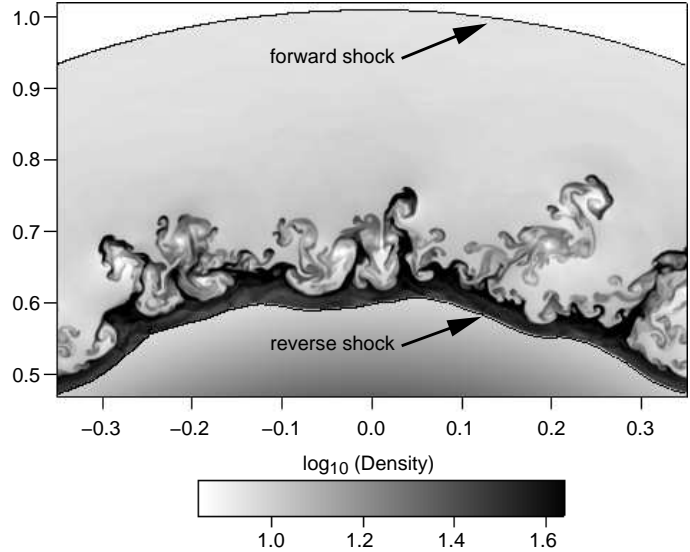}
\end{center}
\caption[]{Two-dimensional calculation of the shock structure for a
supernova with $n=6$ in a stellar wind (courtesy John Blondin).}
\label{fig2}
\end{figure}

 The  density behind the
reverse shock is 
\begin{equation}
\rho_{\rm rev}= 
{(n-4)(n-3)\over 2}
 \rho_{\rm cs}
\label{eq7}
\end{equation}
and is   much higher than behind the
circumstellar shock for $n \gsim 7$. 
There is a drop in density across the contact discontinuity, moving from
the shocked ejecta to the circumstellar medium (see fig. \ref{fig1}).
The fact that low density gas is decelerating higher density gas
leads to a Rayleigh-Taylor instability.
Chevalier, Blondin \& Emmering \cite{CBE92} have calculated the
structure using a two-dimensional PPM (piecewise parabolic method)
hydrodynamic code. They indeed find that
instabilities develop, with dense, shocked ejecta gas penetrating into
the hotter, low density shocked circumstellar gas
(fig. \ref{fig2}). The instability mainly distorts the contact surface,
and does not seriously affect the general dynamics.  The calculation
assumed that cooling is not important. If the gas at the
reverse shock cools efficiently,
the extent of the instability is similar, although the 
Rayleigh-Taylor fingers are narrower \cite{CB95}.

In view of the evidence for dense equatorial winds from red supergiant
stars, Blondin, Lundqvist, \& Chevalier \cite{BLC96} simulated the interaction
of a supernova with such a wind.
They found that for relatively small values of the angular density
gradient, the asymmetry in the interaction shell is greater than,
but close to, that expected from purely radial motion.
If there is an especially low density close to the pole, the flow
qualitatively changes and a protrusion emerges along the axis,
extending to $2-4$ times the radius of the main shell.
Protrusions have been observed in the probable supernova remnant
41.9+58 in M82, although the nature of the explosion is not
clear in this case \cite{Mc01}.

In addition to asymmetric winds, there is evidence for supernova
shock waves interacting with clumps of gas in the wind, as have
been observed in some red supergiant winds (section \ref{mloss}).
In some cases the clumps can be observed by their very narrow
lines in supernova spectra, as in Type IIn supernovae.
The velocity of a shock wave driven into a clump, $v_c$, can be estimated
by approximate pressure balance $v_c\approx v_s (\rho_s/\rho_c)^{1/2}$,
where $v_s$ is the shock velocity in the smooth wind with density
$\rho_s$ and $\rho_c$ is the clump density.
The lower shock velocity and higher density can lead to radiative
cooling of the clump shock although the main shock is non-radiative.
Optical line emission of intermediate velocity observed in Type IIn
(narrow line) supernovae like SN 1978K, SN 1988Z, and SN 1995N
can be explained in this way \cite{CD94,CDD95,Fe01}.

The presence of many clumps can affect the hydrodynamics of the interaction.
Jun, Jones, \& Norman \cite{JJN96} found that propagation in a region
with clumps gives rise to widespread turbulence in the shocked region
between the forward shock and the reverse shock, whereas the
turbulence is confined to a region near the reverse shock for the
non-clump case (fig. \ref{fig2}).
Their simulations are for interaction with a constant density medium,
but the same probably holds true for interaction with a
circumstellar wind.


\section{Emission from the hot gas}

During the first month, radiation from the
supernova photosphere is strong enough for Compton scattering to be the
main cooling process for the circumstellar shock. The photospheric photons
have energies  $\sim 3 ~kT_{eff} \approx 1 - 10$ eV. The optical depth to
electron scattering behind the circumstellar shock is  
\begin{equation}  
\tau_{\rm e} = 0.18 ~\Mdot_{-5} ~u_{\rm w1}^{-1} ~V_4^{-1}~Êt_{\rm days}^{-1}.
\label{eq8}
\end{equation}  
A fraction $\tau_{\rm e}^N$ of the photospheric photons will scatter $N$ times
in the hot gas. In each scattering the photon increases its energy by a  factor
$\Delta\nu/\nu\approx {4 ~k T_{\rm e}/ m_{\rm e} c^2} \gsim 1$.
The multiple scattering creates a power law continuum
that may reach as far up in energy as  the X-ray regime.
If relativistic effects can be ignored ($T_{\rm e} \lsim 10^9 \KK$), the spectral
index is   \cite{F82}
\begin{equation} 
\alpha = \left\{{9\over 4} - {m_{\rm e}c^2\over kT_{\rm e}}~ 
{\rm ln}[~{\tau_{\rm e}\over 2}~
(0.9228 - {\rm ln} ~\tau_{\rm e})]\right\}^{1/2} - {3\over 2}.
\label{eq9}
\end{equation}  
Typically, $1 \lsim \alpha \lsim 3$. 
This type of emission may have been observed in the ultraviolet
emission from SN 1979C \cite{Pe80,F82}.
 For $T_{\rm e} \gsim 10^9 \KK$ relativistic effects become important and
considerably increase the cooling \cite{LF88}.

One can estimate the free-free luminosity from the circumstellar and
reverse shocks from
\begin{equation} 
L_{\rm i} = 4 \pi
\int \Lambda_{\rm ff}(T_{\rm e}) n_{\rm e}^2 r^2 dr \approx \Lambda_{\rm ff}
(T_{\rm i}) {M_{\rm i}
\rho_{\rm i}\over (\mu_{\rm e} m_H)^2 }.
\label{eq10}
\end{equation} 
where the index $i$ refers to quantities connected  either with the reverse shock or circumstellar shock.  The density behind the circumstellar shock is $\rho_{\rm cs} = 
  4 ~\rho_0 = \Mdot /
(\pi u_{\rm w} R_{\rm s}^2) $.  The swept up mass behind the circumstellar 
shock
is $M_{\rm cs}=\Mdot R_{\rm s}/u_{\rm w}$ and that behind the reverse 
shock $M_{\rm rev}= (n-4) M_{\rm cs}/2$. With
$\Lambda_{\rm ff}= 2.4\EE{-27} \bar g_{\rm ff}~ T_{\rm e}^{0.5}$, we get 
\begin{equation} 
L_{\rm i} 
\approx 3.0\EE{39} ~\bar g_{\rm ff}~ C_{\rm n}  ~\left({\Mdot_{-5} \over  
u_{\rm w1} }\right)^2 \left({t \over  
10 \rm ~days}\right)^{-1}~\ergs ,
\label{eq11}
\end{equation}  
where $\bar g_{\rm ff}$ is the free-free Gaunt factor, including relativistic
effects. For the reverse shock
$C_{\rm n} = (n-3) (n-4)^2 / 4 (n-2) $, and for the circumstellar shock $C_{\rm n}=1$.
This assumes electron-ion equipartition, which is highly questionable for
the circumstellar shock (see fig. \ref{fig1}).  
Because of occultation by the ejecta only half of the
above luminosity escapes outward.  

At $T_{\rm e} \lsim 2\EE7 \KK$, line emission increases the cooling rate and
$\Lambda \approx 3.4\EE{-23}~T_{e 7}^{-0.67} \ergs cm^{3}$. If the
temperature of the reverse shock falls below $\sim 2\EE7$ K, a thermal
instability may occur and the gas cools to $\lsim 10^4 \KK$, where
photoelectric heating from the shocks balances the cooling. 
Using $t_{\rm cool} =3kT_{\rm e}/\Lambda$, one obtains for the cooling time
\begin{equation}
t_{\rm cool} =  {605 \over  (n-3) (n-4) (n -
2)^{3.34}}~\left({V_{\rm ej} \over 10^4 \kms}\right)^{5.34} ~\left({\Mdot_{-5}\over
u_{w 1}}\right)^{-1} \left({t\over {\rm days}}\right)^2~~{\rm days},
\label{eq11a}
\end{equation}
assuming solar abundances \cite{FLC96}. From this expression it is clear
that the cooling time is very sensitive to the density gradient, as
well as the shock velocity and mass loss rate. SNe with high mass loss
rates, like SN 1993J, generally have radiative reverse shocks
for $\gsim 100$ days, 
while SNe with lower mass loss rates, like the Type IIP SN
1999em, have adiabatic shocks from early times. 

\begin{figure}[t]
\begin{center}
\includegraphics[scale=0.24,angle=90,origin=c]{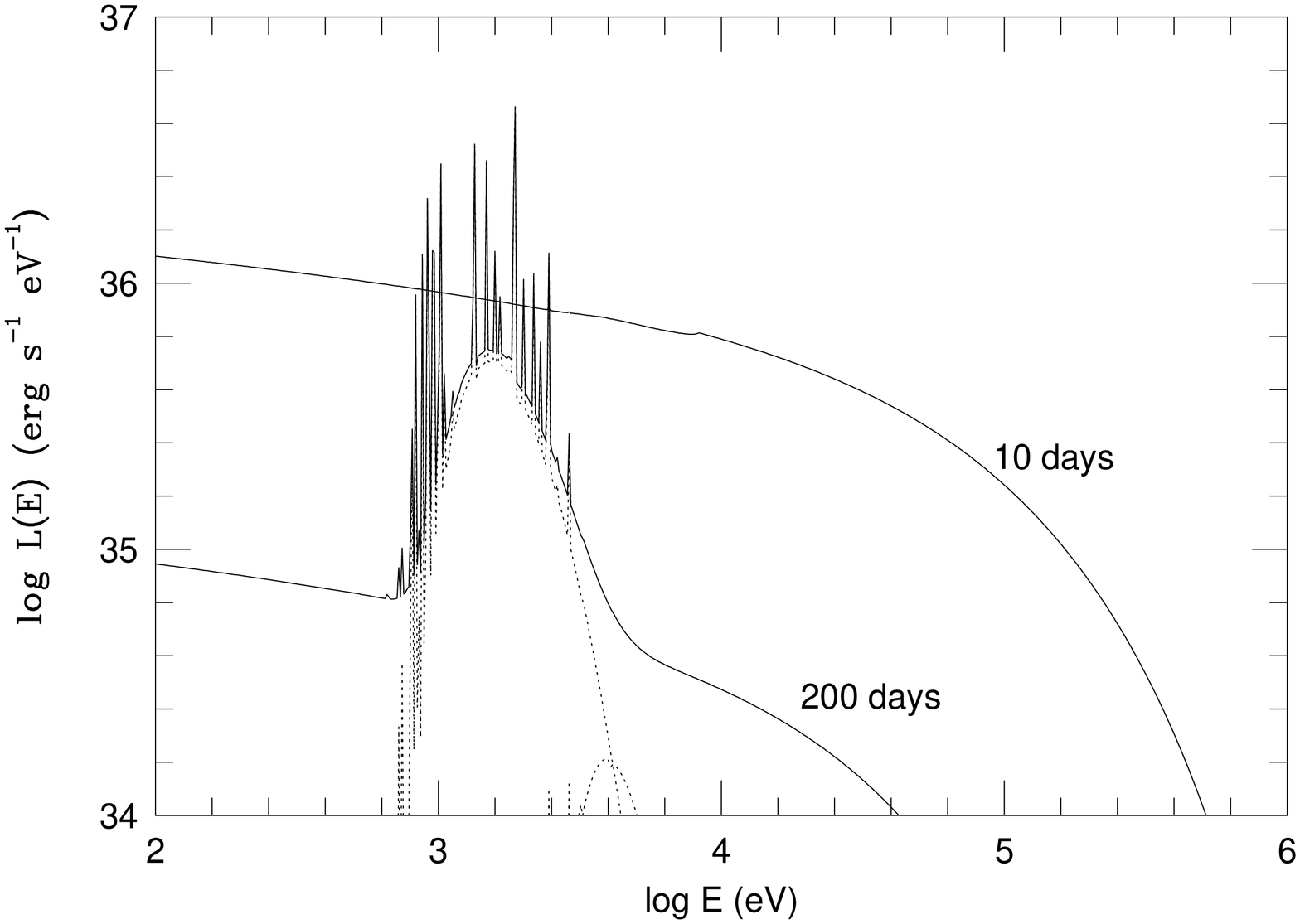}
\includegraphics[scale=0.24,angle=90,origin=c]{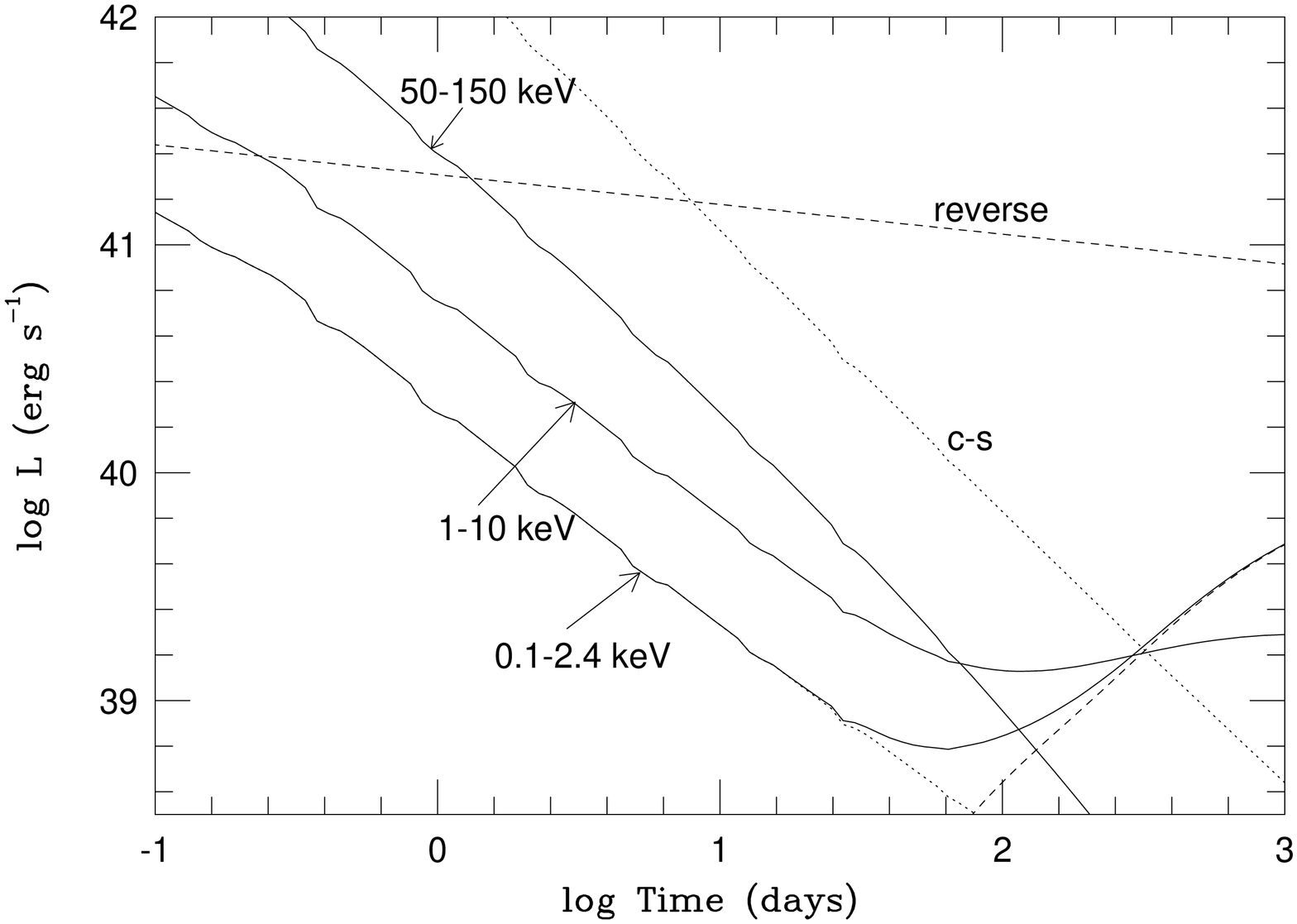}
\end{center}
\caption[]{X-ray spectrum of SN 1993J at 10 days and at 200 days. At
10 days the free-free emission from the outer shock dominates, while
at 200 days the cool shell is transparent enough for the line
dominated spectrum from the reverse shock to dominate instead. Right:
The solid lines give the luminosity in the 0.1-2.4, 1-10, and 50-100
keV bands, corrected for absorption, as a function of time, while the
dotted lines give the total emitted luminosity from the reverse and
circumstellar shocks \cite{FLC96}}
\label{fig3}
\end{figure}

The most important effect of the cooling is that the cool gas may
absorb most of the emission from the reverse shock. Therefore, in
spite of the higher intrinsic luminosity of the reverse shock, little
of this will be directly observable. The column density of the cool gas
is given by $N_{\rm cool} = M_{\rm rev}/(4 \pi R_{\rm s}^2 m_p)$, or
\begin{equation} 
N_{\rm cool} 
\approx 1.0\EE{21} (n-4)  ~\left({\Mdot_{-5} \over  
u_{\rm w1} }\right) \left(V\over 10^4 \kms \right)^{-1} \left({t \over  
100 \rm ~days}\right)^{-1}~\rm cm^{-2}.
\label{eq11b}
\end{equation}
Because the threshold energy due to photoelectric absorption is
related to $N_{\rm cool}$ by $E(\tau=1)=1.2 (N_{\rm cool}/10^{22} ~\rm
cm^{-2})^{3/8}$ keV, it is clear that the emission from the reverse
shock is strongly affected by the cool shell, and a transition from
optically thick to optically thin is expected during the first months,
or year. As an illustration, we show in figure \ref{fig3} the
calculated X-ray spectrum at 10 days and at 200 days for SN 1993J
\cite{FLC96}. At early epochs the spectrum is dominated by the very
hard spectrum from the circumstellar shock, which reaches out to $\gsim
100$ keV. At later epochs the soft spectrum from the reverse shock
penetrates the cool shell, and the line dominated emission from the
cooling gas dominates.  

If cooling, the total energy emitted from the reverse shock is
\begin{eqnarray} 
L_{\rm rev} &=& 4 \pi R_{\rm s}^2 ~{1 \over 2}~ \rho_{\rm ej} V_{\rm rev}^3 = {(n-3)(n-4)\over 4
(n-2)^3}~Ê{\Mdot  V^3\over u_{\rm w} } \cr
&=& 1.6\EE{41} ~ {(n-3)(n-4)\over
(n-2)^3}\Mdot_{-5} u_{\rm w1}^{-1} V_4^3 ~ \ergs.
\label{eq12}
\end{eqnarray}
For high $\Mdot/u_{\rm w}$ the
luminosity from the reverse shock may contribute appreciably, or even dominate, the
bolometric luminosity.

Because $V \propto t^{-1/(n-2)}$, $L_{\rm rev} \propto t^{-3/(n-2)}$ in
the cooling case. Although the total luminosity is  likely to
decrease  in the cooling case, the increasing transparency of the
cool shell, $\tau_{\rm cool} \propto t^{-1}$, can cause the observed flux in
energy bands close to the low energy cutoff, $E(\tau=1)$, to increase
with time, as was seen, e.g., in SN 1993J (figure \ref{fig3}).

Because of the low temperature the spectrum of the reverse shock is
dominated by line emission from metals (fig. \ref{fig3}). An important
point is that the observed spectrum is formed in gas with widely
different temperatures, varying from the reverse shock temperature to
$\sim 10^4$ K. A spectral analysis based on a one temperature model can be
  misleading.  

Chugai \cite{C93} has proposed that the X-ray emission from the
Type IIn SN 1986J is the result of the forward shock front moving
into clumps, as opposed to the reverse shock emission.
One way to distinguish these cases is by the width of line emission;
emission from the reverse shock wave is expected to be broad.
It has not yet been possible to carry through with this test
\cite{He98}.

Another way of observing the hot gas is through the emission from
collisionally heated dust grains in the gas \cite{D87}.
Dust formation in the rapidly expanding ejecta is unlikely,
so the forward shock front must be considered.
Evaporation by the supernova radiation creates a dust-free zone
around the supernova (see section 6.3).
The time for the supernova shock wave to reach the dust depends
on the supernova luminosity and the shock velocity; it is probably
at least several years.
The infrared luminosity from dust can be up to $\sim 100$ times
the X-ray luminosity of the hot gas for typical parameters,
and the dust temperature is a measure of the density of the gas \cite{D87}.
If the X-ray emission from a supernova like
 SN 1986J is from circumstellar clumps
that are out in the region where dust is present, there is the
possibility of a large infrared luminosity.

\section{Radiative heating and re-emission}

\subsection{Soft X-ray burst and circumstellar gas}

The earliest form of circumstellar interaction occurs at shock
break-out. As the shock approaches the surface, radiation leaks out on
a time scale of less than an hour. The color temperature of the
radiation is $\sim (1 - 5)\EE5 \KK$ and the energy $\sim (1-10)\EE{48}
\rm~ergs$ \cite{KI78,F78,EB92,MM99}.
This burst of EUV (extreme ultraviolet)
 and soft X-rays ionizes and heats the
circumstellar medium on a time scale of a few hours. In addition, the
momentum of the radiation may accelerate the circumstellar gas to a
high velocity.  Most of the emission at energies $\gsim 100$ eV is
emitted during the first few hours, and after 24 hours little ionizing
energy remains.

The radiative effects of the soft X-ray burst were most clearly seen from the
ring of SN 1987A, where a number of narrow emission lines from highly
ionized species, like N III-N V, were first seen in the UV
\cite{F89,S9}. Later, a forest of lines came to dominate also the
optical spectrum \cite{W91}. Imaging with HST (e.g., \cite{Je91,Be95}) showed
that the lines originated in the now famous circumstellar ring of SN
1987A at a distance of $\sim 200$ light days from the SN. The presence
of highly ionized gas implied that the gas must have been ionized and
heated by the radiation at shock break-out. 
Because of the finite light travel time across the ring, the
observed total emission from the ring is a convolution of the emission
at different epochs from the various part of the ring. Detailed
modeling \cite{LF91,LF96} shows that while the ionization of the ring
occurs on the time scale of the soft X-ray burst, the gas recombines
and cools on a time scale of years, explaining the persistence of the
emission decades after the explosion. The observed line emission
provides sensitive diagnostics of both the properties of the soft
X-ray burst, and the density, temperature and abundances of the gas in
the ring. In particular, the radiation temperature must have reached
$\sim 10^6$ K, in good agreement with the most detailed recent
modeling of the shock break-out \cite{B99}. 
Narrow emission lines are not unique to SN 1987A, but have also been
observed for several other SNe, in particular several Type IIn SNe,
such as SN 1995N \cite{Fe01} and SN 1998S \cite{Fe00}.

The soft X-ray burst may also pre-accelerate the gas in front of the
shock. In the conservative case that Thompson scattering dominates,
the gas immediately in front of the shock will be accelerated to
\begin{equation}
V = 1.4\EE3~ \left({E \over 10^{48}~ {\rm
ergs}}\right)~
\left({V_{\rm s} \over 1\EE4 \kms}\right)^{-2} ~
 \left({t \over {\rm days }}\right)^{-2}~ \kms, ~
\label{eq16b}
\end{equation}
where
$E$ is the total radiative energy in the burst. Line absorption may
further boost this \cite{F86}. 
If the gas is pre-accelerated, the line
widths are  expected to decrease with time. After about one
expansion time ($\sim R_p/V$) the reverse and circumstellar shocks are
fully developed, and the radiation from these will dominate the
properties of the circumstellar gas. The fraction of this emission
going inward is absorbed by the ejecta and there re-emitted as optical
and UV radiation \cite{F82,F84}.

The X-ray emission from the shocks  ionizes and heats both the
circumstellar medium and the SN ejecta. Observationally, these
components are distinguished easily by the different velocities. The
circumstellar component is expected to have velocities typical of the
progenitor winds, i.e., $\lsim 1000 \kms$, while  the ejecta
have considerably higher velocities. The density is likely to be of the
order of the wind density $10^5-10^7 \ccm$, or higher if clumping is
important.  The ionizing X-ray flux depends strongly on how much of
the flux from the reverse shock  can penetrate the cool shell.
The state of ionization in the circumstellar
gas is characterized by the value of the ionization parameter,
$\zeta=L_{\rm cs}/(r^2 n) = 10^2 (L_{\rm cs}/10^{40} \ergs) (r/10^{16} \rm~
cm)^{-2} (n / 10^6 \ccm)^{-1}$ \cite{KMC82}. The
comparatively high value of $\zeta \approx 10-10^3$ explains the
presence of narrow coronal lines of [Fe V-XI] seen in objects like SN 1995N
\cite{Fe01}.

\subsection{SN ejecta}
\begin{figure}[t]
\begin{minipage}[c]{0.5\textwidth}
\centering 
\includegraphics[scale=0.3]{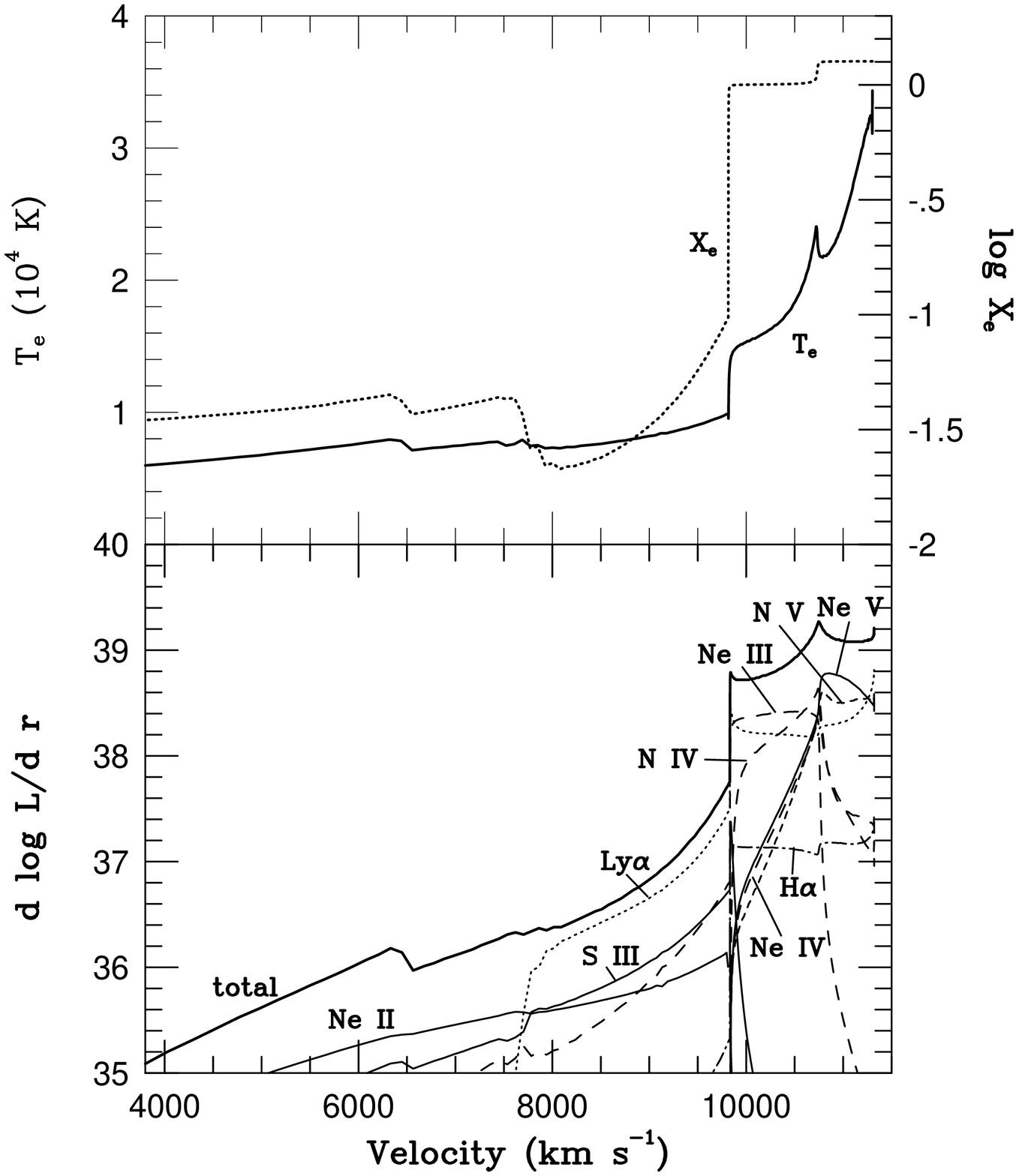}
\end{minipage}%
\begin{minipage}[c]{0.5\textwidth}
\centering 
\includegraphics[scale=0.3]{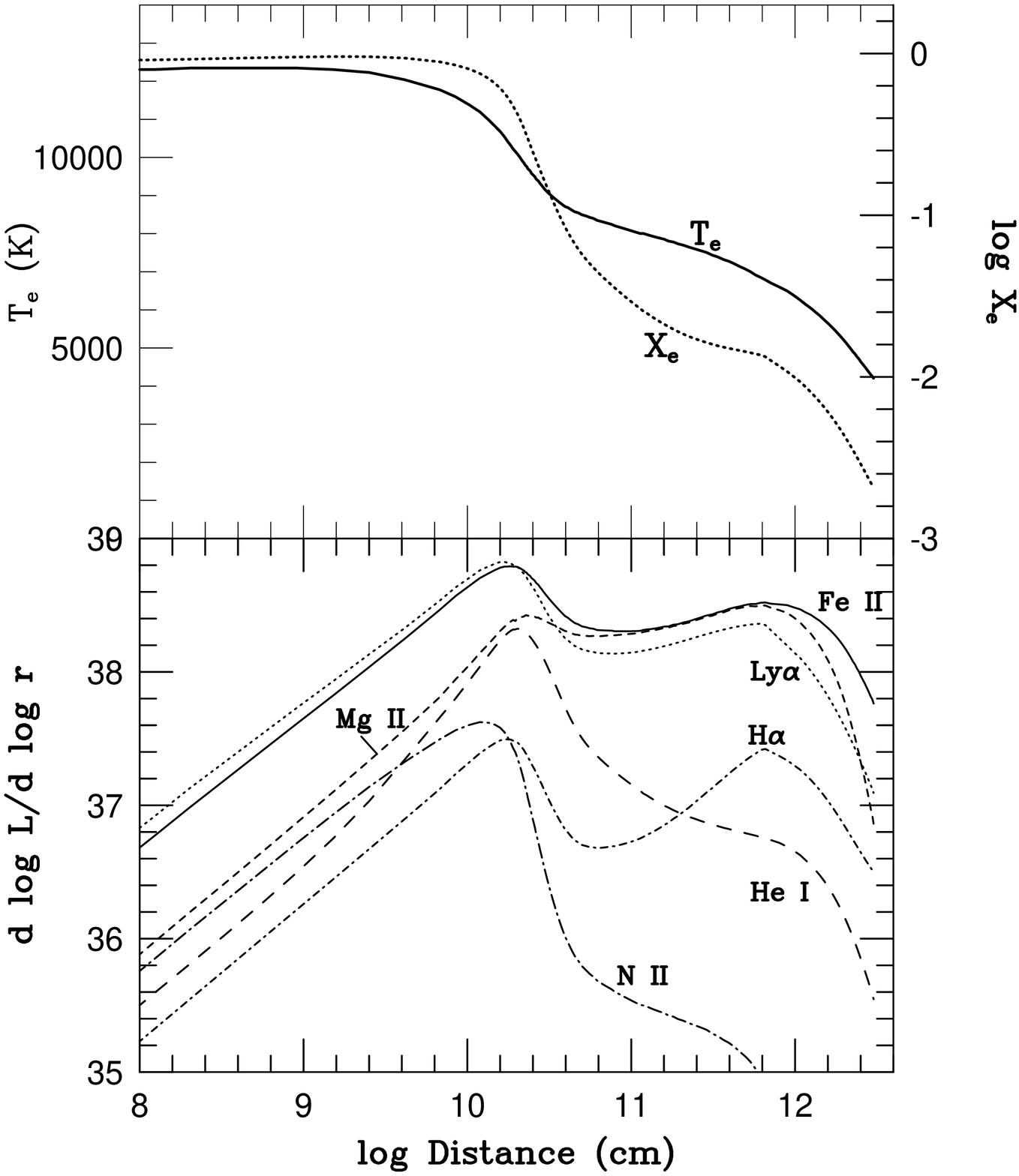}
\end{minipage}
\caption[]{Structure of the ejecta and cool shell ionized by the
reverse shock at 500 days for parameters appropriate to SN 1993J ($\Mdot = 5\EE{-5} \Msunyr$ for $u = 10 \kms$). Upper
panels show the temperature and ionization of the ejecta (left panels)
and the cool shell (right panels), while the lower panels show the
corresponding luminosities per unit distance. Note the different
length scales in the two panels. The ejecta region has low density,
high temperature and ionization, while the cool shell has a high
density, is extremely thin, has a low temperature, and is only
partially ionized.}
\label{fig5}
\end{figure}

The ingoing X-ray flux from the reverse shock  ionizes the outer
parts of the ejecta. The state of highest ionization  therefore is
close to the shock, with a gradually lower degree of ionization
inwards. Unless clumping in the ejecta is important, the ejecta
density is $\sim 10^6-10^8 \ccm$. In the left panel of figure
\ref{fig5} we show temperature and ionization structure of the ejecta,
as well as the emissivity of the most important lines. The temperature
 close to the shock is $\sim 3\EE4$ K. Calculations show that most of
the emission here is  emitted as UV lines of highly ionized ions, like
\La, C III-IV, N III-V, and O III-VI. Inside the ionized shell
there is an extended partially ionized zone, similar to that
present in the broad emission line regions of AGN's. Most of the
emission here comes from Balmer lines.

As we have already discussed, the outgoing flux from the reverse shock
is to a large extent absorbed by the cool shell between reverse shock
and the contact discontinuity if radiative cooling has
been important. The whole region behind the reverse
shock is in approximate pressure balance, and the density of this gas
is therefore be a factor $\sim 4 T_{\rm rev}/T_{\rm cool} \approx 10^3-10^4$
higher than that of the ejecta. Because of the high density, the gas
is only be partially ionized and the temperature only $(5-8)\EE3$ K. 
Most of the emission comes out as Balmer lines, Mg II and Fe II
lines (figure \ref{fig5}, right panel). The thickness of the emitting
region is also very small, $\sim 3\times 10^{12}$ cm. In one
dimensional models, the velocity is marginally smaller than the
highest ejecta velocities. Instabilities in the shock are, however,
likely to erase this difference.

An important diagnostic of the emission from the cool shell and the
ejecta is the \Ha~ line. This line arises as a result of recombination
and collisional excitation. In \cite{CF94}, it is shown that $\sim 1 \%$ of
the reverse shock luminosity is emitted as \Ha, fairly independent of
density and other parameters. Observations of this line 
permit us to follow the total luminosity from the reverse shock,
complementary to the X-ray observations. 
In SN 1993J, the \Ha~ line had the box-like shape that is expected
for shocked, cooled ejecta \cite{M00a,M00b}.
The top of the line showed structure that varied with time; this
could be related to hydrodynamic instabilities of the reverse
shocked gas.

\subsection{Interaction with dust}

The winds from red supergiant stars are known to contain dust, so that
infrared emission from radiatively heated dust and dust scattering
of supernova light might be expected.
The dust temperature is determined by a balance between radiative
heating and emission from the grain surface.
If the supernova radiates like a blackbody of temperature $T_{\rm sn}$
and if the dust absorption efficiency varies as $\lambda^{-q}$,
the grain temperature is $T_{\rm g}=T_{\rm sn}W^{1/(4+q)}$, where
$W$ is the dilution factor for the supernova radiation \cite{D83}.
For $q=1$, we have
\begin{equation}
T_{\rm g}= 280 \left(T_{\rm sn}\over 5000{\rm~K}\right)^{0.2}
\left(L_{\rm sn}\over 10^{42} {\rm~ergs~s^{-1}}\right)^{0.2}
\left(r\over 10^{18} {\rm~cm}\right)^{-0.4}  {\rm~ K},
\end{equation}
where $L_{\rm sn}$ is the supernova luminosity.

Because typical evaporation temperatures for grain materials are in the range
$1000-1500$ K, the dust near the supernova is evaporated.
The radius out to which dust is evaporated, $r_v$, is probably determined
by the luminosity at the time of shock breakout, when $L_{\rm sn}$
can be $\gsim 10^{44}$ ergs s$^{-1}$.
SN 1979C and SN 1980K were detected as infrared sources, and Dwek \cite{D83}
estimates $r_v\approx 3\times 10^{17}$ cm in these cases.
The infrared light curve can be calculated from $T_{\rm g}$, taking
into account light travel time effects.
If the characteristic time that the supernova is bright is short
compared to $2r_v/c$, there is a plateau phase until a time $2r_v/c$.
The infrared flux then drops as $t^{-2}$ for a highly extended wind
and drops more rapidly if the cutoff in the wind, $r_w$, is close to $r_v$.
For SN 1979C and SN 1980K, Dwek \cite{D83} found $r_w\lsim 10^{18}$ cm.
The ratio of the infrared emitted energy to the total emitted energy
gives an estimate of the optical depth through the dusty wind.
The optical depths for SN 1979C and SN 1980K are $\sim 0.3$ and 0.03,
respectively, leading to minimum shell masses of $\sim 1-5~M_{\odot}$ and
$\sim 0.1-0.4~M_{\odot}$ for the two supernovae \cite{D83}.
The corresponding mass loss rates are consistent with those derived
from radio observations \cite{LF88}.
Infrared emission that can
be attributed to radiatively heated dust 
has recently been observed from SN 1998S \cite{Fe00}.
Infrared dust echoes have the potential to give information on
dust composition and distribution \cite{D85,EC88}, but the observations
do not yet exist to address these issues.

The dust grains that give an infrared echo can also give a scattered
light echo \cite{C86}.
The ratio between the total scattered light and the infrared light
depends on the albedo of the dust.
The light curve decreases with time before $2r_v/c$ in this case because
of the strong forward scattering of typical grains.
Chevalier \cite{C86} failed to find good evidence for scattered light
echoes from SN 1979C and SN 1980K, although such echoes might have
been observable.
The implication may be that the dust grains have small albedos.
The problem with detecting scattered light echoes is that
any echo light may be dominated by light coming directly from
the circumstellar interaction.
Roscherr \& Schaefer \cite{RS00} examined the late emission from
two Type IIn supernovae and determined that it was due to shock interaction.
Because of the ability to spatially resolve the emission in SN 1987A,
it has been possible to observe scattered light from the wind
interaction nebula \cite{CKH95} and possibly a shell at the outer
extent of the red supergiant wind \cite{CE89}.

\section{Relativistic Particles}

Unambiguous evidence for the presence of relativistic electrons comes from radio
observations of SNe. A characteristic is the wavelength-dependent
turn-on of the radio emission (\cite{SW90}; chapter by Sramek), first
seen at short wavelengths, and later at longer wavelengths. 
This behavior is interpreted  as a result of 
decreasing absorption due to the expanding emitting region \cite{C82b}.

Depending on the magnetic field and the density of the circumstellar
medium, the absorption may be produced either by free-free absorption
in the surrounding thermal gas, or by synchrotron self-absorption by
the same electrons that are responsible for the emission. The
relativistic electrons are believed to be produced close to the
interaction region, which provides an ideal environment for the
acceleration of relativistic particles. The details of the
acceleration and injection efficiency are still not well understood
(see, e.g., \cite{DM93} and references therein). 
 Here we just parameterize the
injection spectrum with the power law index $p_{\rm i}$ and an
efficiency, $\eta$, in terms of the  postshock energy density. Without
radiation or collisional losses the spectral index of the synchrotron
emission will then be $\alpha=(p-1)/2$, where flux $\propto \nu^{-{\alpha}}$.
 Diffusive acceleration
predicts that $p_{\rm i} = 2$ in the test particle limit.
If the particle acceleration is very efficient and nonlinear effects
are important, the electron spectrum can be steeper
(e.g., \cite{ER91}).

For free-free absorption, the optical depth $\tau_{\rm ff} =
\int_{R_{\rm s}}^{\infty} \kappa_{\rm ff} n_{\rm e} n_{\rm i} dr$
 from the radio emitting
region close to the shock through the circumstellar medium decreases as
the shock wave expands,
explaining the radio turn-on.  Assuming a fully ionized wind with
constant mass loss rate and velocity, so that equation (\ref{eq1b})
applies, the free-free optical depth at wavelength $\lambda$ is
\begin{equation} 
\tau_{\rm ff}(\lambda) \approx 7.1\times 10^{2}
\lambda^2\left({\Mdot_{-5}\over u_{w 1}}\right)^2 T_5^{-3/2} V_4^{-3}
t_{\rm days}^{-3}
\label{eq13}
\end{equation}
where $\Mdot_{-5}$ is the mass loss rate in units of $10^{-5}
~\Msunyr$, $u_{w 1}$ the wind velocity in units of $10 \kms$, and
$T_5$ the temperature of the circumstellar gas in $10^5 \KK$.  From
the radio light curve, or spectrum, the epoch of $\tau_{\rm ff} = 1$ can
be estimated for a given wavelength, and from the line widths in the
optical spectrum the maximum expansion velocity, $V$, can be
obtained. Because the effects of the radiation from the supernova have
to be estimated from models of the circumstellar medium, the
temperature in the gas is uncertain. Calculations show
that initially the radiation heats the gas to $T_{\rm e} 
\approx 10^5 \KK$ \cite{LF88}.
 $T_{\rm e}$ then decreases with time, and after a year $T_{\rm e} \approx
(1.5-3)\EE4 \KK$. In addition, the medium may recombine, which further
decreases the free-free absorption. From $t[\tau(\lambda)_{\rm ff}=1]$ the
ratio  $\Mdot/u_{\rm w}$ can be calculated. Because $\Mdot/u_{\rm w} \propto
T_{\rm e}^{3/4} x_{\rm e}^{-1}$, errors in $T_{\rm e}$ and $x_{\rm e}$ may 
lead to large
errors in $\Mdot$. If the medium is clumpy, equation (\ref{eq13})
 may lead to an
overestimate of $\Mdot/u_{\rm w}$.

Under special circumstances (see below), synchrotron self-absorption
(SSA) by
the same relativistic electrons emitting the synchrotron radiation may
be important \cite{S90,C98,FB98}. 
The emissivity of the synchrotron plasma is given by
  $j(\lambda)= {\rm const.}~ \lambda^{\alpha} B^{1+\alpha}
N_{\rm rel}$ \cite{L92},
 while the optical depth to self-absorption is given by
\begin{equation}
\tau_{\rm s} = {\rm const.} ~\lambda^{(5/2)+\alpha} B^{(3/2)+\alpha} N_{\rm rel}~.
\label{eq13a}
\end{equation}
Here $N_{\rm rel}$ is the column density of relativistic electrons and
$B$ the magnetic field. The flux from a disk with radius $R_{\rm s}$ of
relativistic electrons is, including the effect of SSA, given by
\begin{equation} 
F_\nu(\lambda) \propto R_{\rm s}^2 S(\lambda)[1-e^{-\tau_{\rm
s}(\lambda)}],
\label{eq13b}
\end{equation}
where $S(\lambda) = j(\lambda)/\kappa(\lambda) = {\rm
const.}~\lambda^{-5/2} B^{-1/2}$ is the source function. In the
optically thick limit we therefore have $F_\nu(\lambda) \approx {\rm
const.} ~R_{\rm s}^2~ \lambda^{-5/2} B^{-1/2}$, independent of $N_{\rm rel}$. A
fit of this part of the spectrum therefore gives the quantity $R_{\rm s}^2~
B^{-1/2}$. The break of the spectrum determines the wavelength of
optical depth unity, $\lambda(\tau_{\rm s}=1)$. Equation (\ref{eq13a})
therefore gives a second condition on $B^{3/2+\alpha}~N_{\rm
rel}$. If $R_{\rm s}(t)$ is known in some independent way, one can
 determine both the magnetic field and the column density of
relativistic electrons, independent of assumptions about equipartition,
etc. 
In some cases, most notably for
SN 1993J, the shock radius, $R_{\rm s}$, can be determined directly from VLBI
observations. If this is not possible, an alternative is from
observations of the maximum ejecta velocity seen in, e.g., the \Ha
~line, which should reflect the velocity of the gas close to the
shock.  Because the SN expands homologously, $R_{\rm s} = V_{\rm max} t$. A fit
of the spectrum at a given epoch can therefore yield both $B$ and
$N_{\rm rel}$ independently. From observations at several epochs the
evolution of these quantities can then be determined.

Although the injected electron spectrum from the shock is likely to be
a power law with $p_{\rm i} \approx 2$ ($\alpha \approx 0.5$), the
integrated electron spectrum is affected by various loss
processes. Most important, the synchrotron cooling time scale of an
electron with Lorentz factor $\gamma$ is $t_{\rm syn} \approx
9\EE{3} \gamma^{-1} B^{-2}$ days. This  especially affects 
the high energy electrons, steepening the index of the electron
spectrum by one unit, $p =p_{\rm i}+1 \approx 3$ ($\alpha \approx
1$). Inverse Compton losses have a similar effect. At low energy,
Coulomb losses may be important, causing the electron spectrum to
flatten.

The best radio observations of any SN were obtained for SN
1993J. This SN was observed from the very beginning until late epochs
with the VLA at wavelengths between 1.3 -- 90 cm \cite{vD94}, producing a
set of beautiful light curves (see also Chapter by Sramek). In
addition, the SN was observed  with VLBI \cite{M95,M97,B00},
resulting in an impressive sequence of images 
in which the radio emitting plasma could be
directly observed. These images showed a remarkable degree of symmetry
and clearly resolved the shell of emitting electrons. The evolution of
the radius of the radio emitting shell could be well fitted by $R_{\rm s}
\propto t^{0.86}$, implying a deceleration of the shock front. 

\begin{figure}[t]
\begin{center}
\includegraphics[scale=0.32,origin=c]{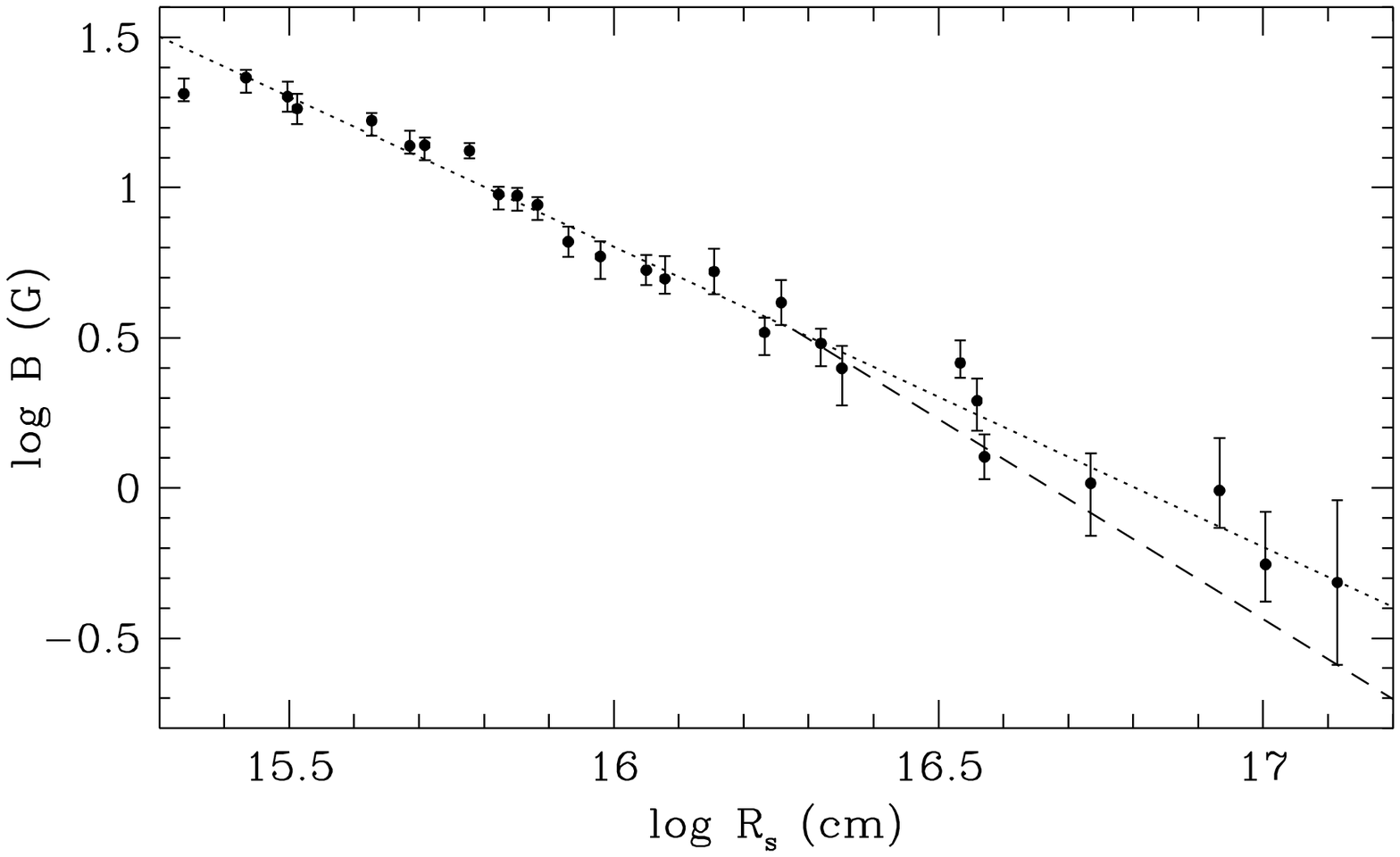}
\includegraphics[scale=0.32,origin=c]{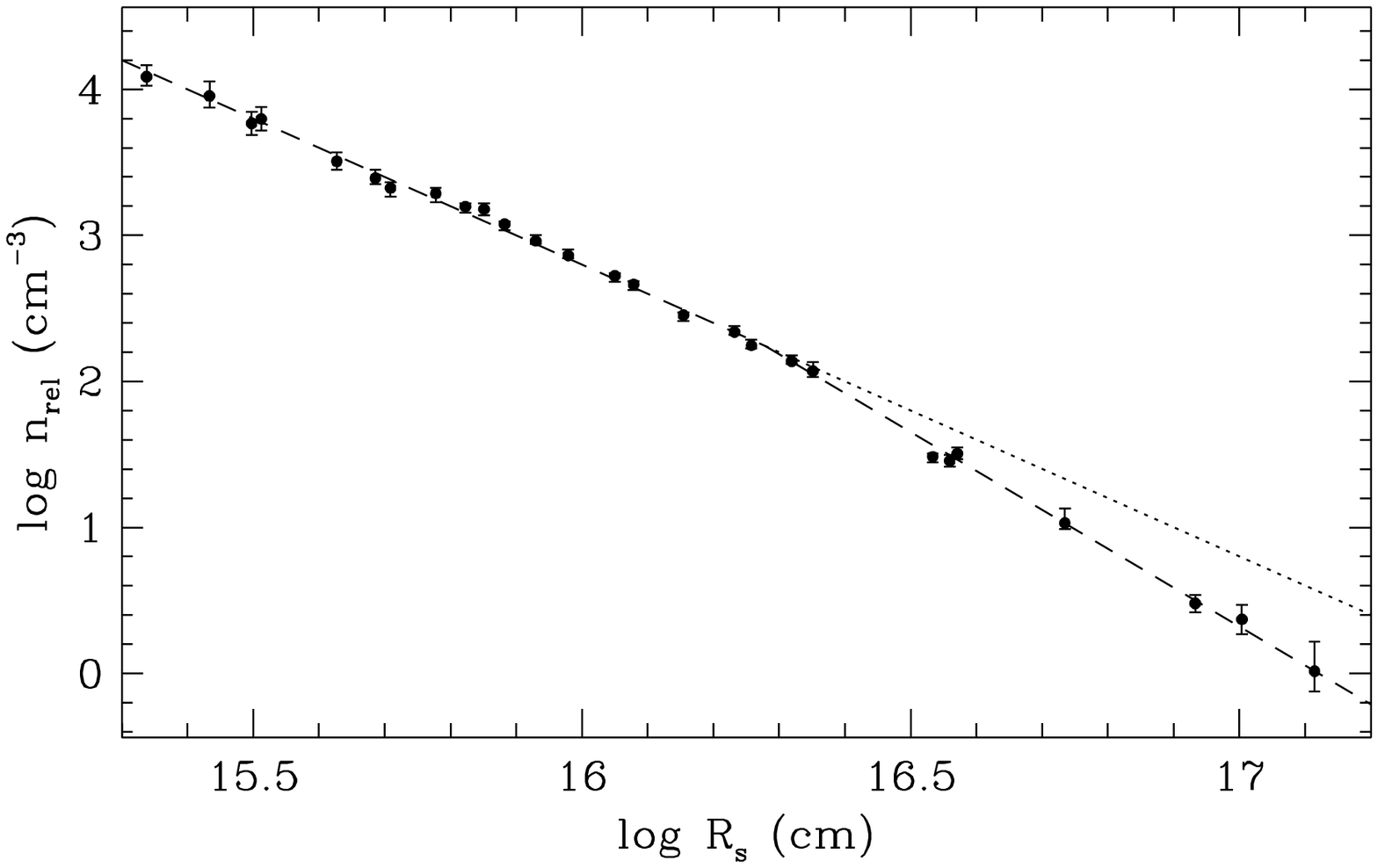}
\end{center}
\caption[]{Magnetic field (left) and density of relativistic electrons
(right) as a function of the shock radius for SN 1993J. The dashed
lines show the expected evolution if the magnetic energy density and
relativistic particle density scale with the thermal energy density,
$B^2/8 \pi \propto \rho V_{\rm s}^2 \propto n_{\rm rel} \propto
t^{-2}$, while the dotted lines show the case when $B \propto r^{-1}$
and $n_{\rm rel} \propto r^{-2}$\cite{FB98}.}
\label{fig4}
\end{figure}
From a fit of the observed spectra for the different epochs the
magnetic field and number of relativistic electrons could be
determined for each epoch, as described above \cite{FB98}. In figure
\ref{fig4} we show the evolution of $B$ and particle density
$n_{\rm rel}$, plotted as
a
function of the shock radius. The most remarkable thing is the smooth
evolution of these quantities, showing that $B \approx 6.4 (R_{\rm
s}/10^{16} ~{\rm cm})^{-1}$ G, and $n_{\rm rel} \propto \rho V^2
\propto t^{-2}$, the thermal energy density behind the shock. The
magnetic field is close to equipartition, $B^2/8 \pi \approx 0.14 ~\rho
V_{\rm s}^2 $, much higher than expected if the circumstellar magnetic
field, of the order of a few mG, was just compressed, and strongly
argues for field amplification, similar to what has been seen in
simulations \cite{JN96}. Contrary to earlier, simplified models for SN 1993J 
based
on free-free absorption only \cite{FLC96,vD94}, the circumstellar
density was consistent with $\rho \propto r^{-2}$.

\begin{figure}[t]
\centering
\includegraphics[scale=0.30,angle=-90,origin=c]{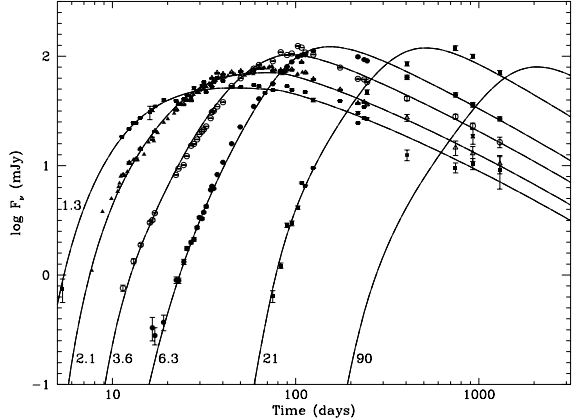}
\caption[]{Observed \cite{vD94} and model radio light curves of SN 1993J \cite{FB98}.}
\label{fig4a}
\end{figure}

In figure \ref{fig4a}, we show the excellent fit of the resulting light
curves. The high values of $B$ implied that synchrotron cooling was
important throughout most of the evolution for the electrons
responsible for the cm emission, and also for
the 21 cm emission before $\sim 100$ days.
 At early epochs, Coulomb losses were important for
the low energy electrons. The injected electron spectrum was best
fitted with $p_{\rm i} = 2.1$.

The form of the light curves can be understood if, for simplicity, we
{\it assume} equipartition, so that $B^2/8 \pi = \eta \rho V_{\rm
s}^2$. With $\rho \propto (\Mdot/u)~R_{\rm s}^{-2}$ and $V_{\rm s}
\propto R_{\rm s}/t$, we find that $B \propto (\Mdot/u)^{1/2} t^{-1}$. The
optically thick part is therefore given by
\begin{equation} 
F_\nu(\lambda) \propto ~R_{\rm s}^2~ \lambda^{-5/2} B^{-1/2} 
\propto(\Mdot/u)^{-1/4} \lambda^{-5/2}  t^{(5n-14)/2(n-2)},
\label{eq13c}
\end{equation}
since $R_{\rm s} \propto t^{(n-3)/(n-2)}$. For large $n$, we get
$F_\nu(\lambda) \propto t^{5/2}$.  An additional curvature of the spectrum
is produced by free-free absorption in the wind, although this only
affects the spectrum at early epochs.

In the optically thin limit, $F_\nu(\lambda) \propto R_{\rm s}^2
j(\lambda)\propto R_{\rm s}^2 \lambda^{\alpha} B^{1+\alpha} N_{\rm rel}$. If
losses are unimportant, $N_{\rm rel, tot} = 4 \pi R_{\rm s}^2 N_{\rm rel}$,
the total number of relativistic electron, may either be assumed to be
proportional to the total mass, if a fixed fraction of the shocked
electrons are accelerated, or be proportional to the swept up thermal
energy. In the first case, $N_{\rm rel, tot} \propto \Mdot R_{\rm
s}/u_w$, while in the second $N_{\rm rel, tot} \propto \Mdot R_{\rm s}
V_{\rm s}^2 /u_w$, so that in general $N_{\rm rel, tot} \propto \Mdot
R_{\rm s} V_{\rm s}^{2\epsilon} /u_w$, where $\epsilon = 0$ or $1$ in
these two cases. Therefore, $F_\nu(\lambda) \propto \Mdot/u ~ R_{\rm
s} V_{\rm s}^{2 \epsilon} \lambda^{\alpha} B^{1+\alpha}$. If the
B-field is in equipartition, as above, and using $V = (n-3)/(n-2) R_{\rm s}/t
\propto t^{-1/(n-2)}$ we find 
\begin{equation} 
F_\nu(\lambda) 
\propto (\Mdot/ u_w)^{(3+\alpha)/2} ~\lambda^{\alpha} t^{-\alpha -
(1+2\epsilon)/(n-2)}.
\label{eq13d}
\end{equation}
If synchrotron cooling is important, a similar
type of expression can be derived \cite{FB98}. The main thing to note
is, however, that the optically thin emission is expected to be
proportional to the mass loss rate, $\Mdot/u_w$, and that the decline rate
depends on whether the number of relativistic particles scale with the
number density or the thermal energy of the shocked gas, as well as
spectral index. Observations
of the decline rate can therefore test these possibilities.

Although a self-consistent model can be developed for SN 1993J and
other radio supernovae, the modeling of SN 1986J and related objects
has been unclear.
Weiler, Panagia, \& Sramek \cite{WPS90} proposed a model for SN 1986J
in which thermal absorbing material is mixed in with the nonthermal
emission; one possibility for this is a very irregular shocked
emission region.
The absorption is described in a parameterized way and is not related
to a quantitative supernova model.
Chugai \& Belous \cite{CB99} propose a model in which the absorption
is by clumps.
The narrow line optical emission implies the presence of clumps,
but they are different from those required for the radio absorption.
The possible presence of clumps and irregularities introduces
uncertainties into models for the radio emission, although rough
estimates of the circumstellar density can still be obtained.

\begin{figure}[t]
\centering
\includegraphics[viewport=  0 10 600 600,scale=0.40,angle=0,origin=c,clip]
{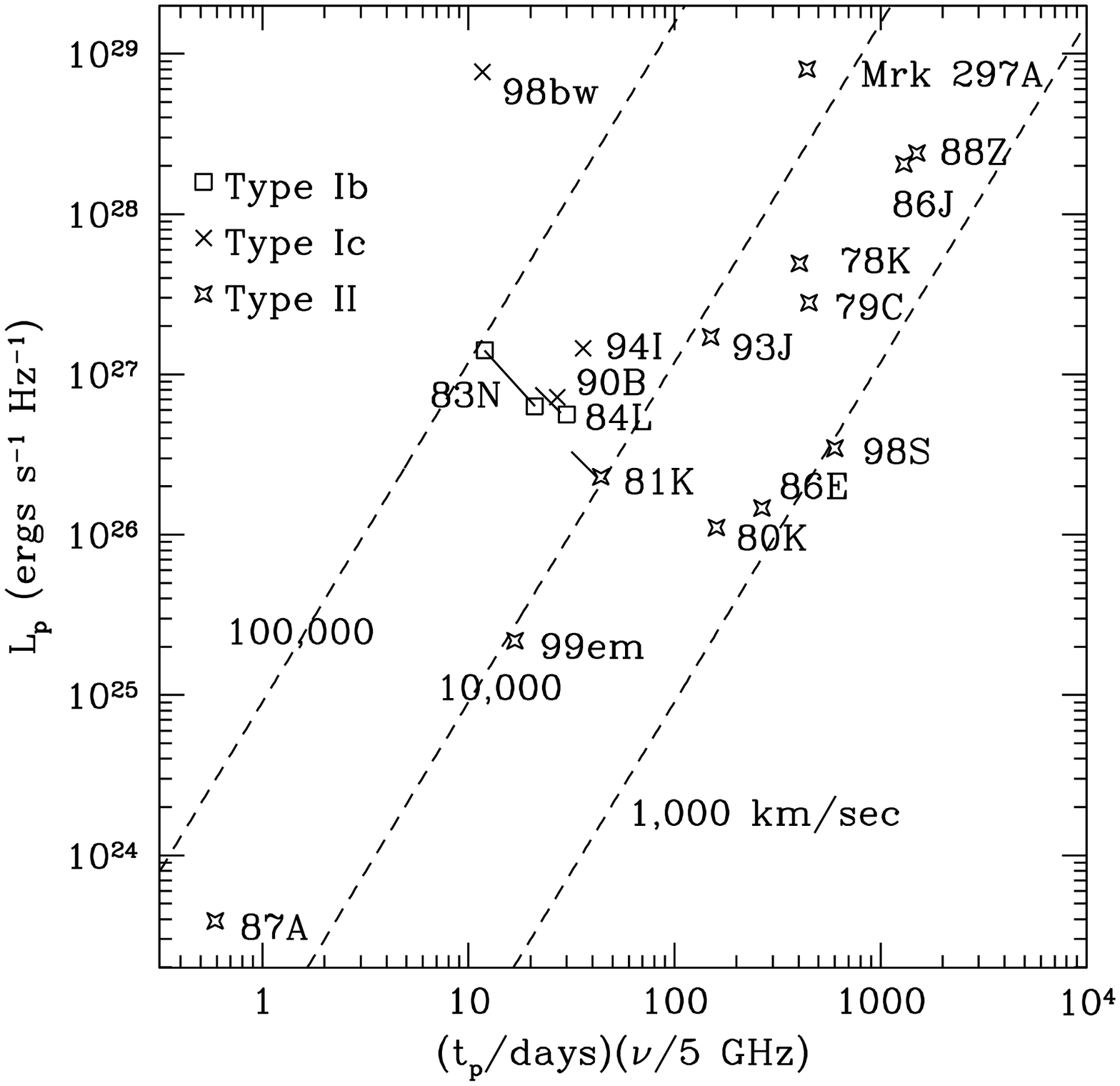}
\caption[]{Peak luminosity and corresponding epoch for the
well-observed radio SNe. The dashed lines give curves of constant
expansion velocity, {\it assuming} SSA (see \cite{C98}).}
\label{fig4d}
\end{figure}
The relative importance of SSA and free-free absorption depends on a
number of parameters and we refer to \cite{C98,FB98} for a more
detailed discussion. The most important of these are the mass loss
rate, $\Mdot/u_w$, the shock velocity, $V_{\rm s}$, and the
circumstellar temperature, $T_{\rm e}$. In general, a high shock
velocity and a high circumstellar temperature favor SSA, while a high
mass loss rate favors free-free absorption.  

Assuming SSA, an interesting expression for the velocity of the shock
can be derived which can be tested against the observations. If we 
 assume equipartition, $N_{\rm rel} \propto \rho  
V_{\rm s}^{2} R_{\rm s}$ and $B^2/8 \pi \propto \rho V_{\rm s}^2$ we
have $N_{\rm rel} \propto B^2 R_{\rm s}$. Using this in equation
(\ref{eq13a}) we get $\tau_{\rm s} = {\rm const} ~\lambda^{5/2+\alpha}
B^{7/2+\alpha} ~R_{\rm s}$. The peak in the light curve is given by
$\tau_{\rm s} \approx 1$. If we approximate the flux at this point by
the optically thick expression equation (\ref{eq13c}) and solve for
$B$ we get $B \propto F_\nu(\lambda)^{-2} ~R_{\rm s}^4~
\lambda^{-5}$. Inserting this expression in the condition $\tau_{\rm s}
\approx 1$, we find $R_{\rm s}^{15+4\alpha}
F_\nu(\lambda)^{-7-2\alpha} ~ \lambda^{-15-4\alpha} \approx {\rm const}$. With
$V_{\rm s} = (n-3)/(n-2) R_{\rm s}/ t$ we finally have
\begin{equation} 
 V_{\rm s} \approx {\rm const}~ F_\nu(\lambda)^{(7+2\alpha)/(15+4\alpha)} ~
\lambda~ t^{-1}
\label{eq13e}
\end{equation}
where all parameters refer to their values at the peak of the light
curve.  Using this expression, we can plot lines of constant shock
velocity into a diagram with peak radio luminosity versus time of peak
flux, {\it assuming} that SSA dominates (figure \ref{fig4d}).
The positions of the lines depend only weakly on the equipartition
assumption. Each SN
can now be placed in this diagram to give a predicted shock
velocity. If this is lower than the observed value (as measured by
VLBI or from line profiles) SSA gives a too low flux and should
therefore be relatively unimportant and free-free absorption instead
dominate. In figure \ref{fig4d} we show an updated version of the
figure in \cite{C98}. The most interesting point is that most Type
Ib/Ic SNe, SN 1983N, SN 1994I and SN 1998bw, fall into the high
velocity category, while Type IIL SNe, like SN 1979C and SN 1980K, as
well as the Type IIn's SN 1978K, SN 1988Z, SN 1998S fall in the
free-free group. SN 1987A is clearly special with its low mass loss
rate, but is most likely dominated by SSA \cite{SM87,S90,CD95}.
Mrk 297A is an apparent radio supernova in the clumpy, irregular
galaxy Mrk 297 \cite{Y94}; its supernova type is not known.

\section{Discussion and conclusions}

Circumstellar interaction of supernovae gives an important window on
the nature of stars that explode and their evolution leading up to
the explosion.
Mass loss rates for the red supergiant progenitors of Type II supernovae
range from $\sim 2\times 10^{-6}~\Msunyr$ for SN 1999em \cite{Pe01}
to $\gsim 2\times 10^{-4}~\Msunyr$ for SN 1979C and SN 1986J
\cite{LF88}.
Evidence for CNO processing has been found in a number of supernovae,
including SN 1979C \cite{Fe84}, SN 1987A \cite{F89} and SN 1995N \cite{Fe01}.
In some cases, the reverse shock appears to be moving into gas that
is H poor and O rich, e.g., SN 1995N \cite{Fe01}; this relates to
the total amount of mass loss before the supernova.
The complex circumstellar environment of SN 1987A has become clear
because of it proximity (see Chapter by McCray).
For more distant supernovae, studies of polarization and spectral
line profiles can reveal asymmetries, as in SN 1998S \cite{Le00}.

The evidence on circumstellar interaction is especially useful
when it can be combined with information from other aspects of
the supernovae, such as their light curves and stellar environments.
For example, from the pre-supernova stellar environment of 
SN 1999em, Smartt et al. \cite{Sme01} deduced an initial mass
of $12\pm 1~M_{\odot}$.
The supernova was of the plateau type, implying that hydrogen
envelope was largely intact at the time of the supernova.
This is consistent with the relatively low rate of mass loss
deduced for the supernova progenitor \cite{Pe01}.

In addition to information on the evolution of massive stars and their
explosions, circumstellar interaction provides an excellent laboratory
for the study of shock wave physics.  Compared to older supernova
remnants, the shock velocities are higher and the time evolution gives
an additional dimension for study, although there is little spatial
information in most cases. VLBI observations can, however, in this
respect be extremely valuable, as demonstrated by SN 1993J.  An
object where both the spatial and time dimensions are accessible is SN
1987A, which has turned out to be an excellent source for the study of
shock waves (see Chapter by McCray).

We are grateful to John Blondin for providing Figure 2 and to Peter Lundqvist for comments on the manuscript.
This work was supported in part by NASA grant NAG5-8232 and by the Swedish Research Council.

%

%

\end{document}